\documentclass[fleqn]{article}
\usepackage{longtable}
\usepackage{graphicx}

\begin {document}
\begin{flushleft}
{\LARGE
{\bf Comment on ``Atomic data for Ne-like ions useful in plasma diagnostic" by  Singh et al.  [Can. J. Phys.   96 (2018) 36]}
}\\

\vspace{1.5 cm}

{\bf {Kanti  M  ~Aggarwal}}\\ 

\vspace*{1.0cm}

Astrophysics Research Centre, School of Mathematics and Physics, Queen's University Belfast, \\Belfast BT7 1NN, Northern Ireland, UK\\ 
\vspace*{0.5 cm} 

e-mail: K.Aggarwal@qub.ac.uk \\

\vspace*{0.20cm}


\vspace*{1.0 cm}


\vspace*{1.0 cm}
\vspace*{1.0 cm}

\hrule

\vspace{0.5 cm}

\end{flushleft}

\clearpage


\begin{abstract}

In a recent paper, Singh et al. [Can. J. Phys.  {\bf 96} (2018) 36] have reported results for  energy levels, radiative rates, and lifetimes  among 209 levels of four Ne-like ions, namely Hf~LXIII, Ta~LXIV, W~LXV, and Re~LXVI. For their calculations, they have adopted the GRASP and FAC codes, and have assessed their energy levels to be accurate to $\sim$0.5~Ryd,  based on comparisons between the two sets of energies. However,  some of the levels between the two calculations differ by up to $\sim$2~Ryd, and for three ions. In addition, we also note that some of their results with FAC cannot be reproduced, and hence the large discrepancies.

\end{abstract}

\clearpage

\section{Introduction}

In a recent paper, Singh et al. \cite{sam} have reported results for  energy levels, radiative rates, and lifetimes  among 209 levels of four Ne-like ions, namely Hf~LXIII, Ta~LXIV, W~LXV, and Re~LXVI. These levels belong to the 2s$^2$2p$^6$, 2s$^2$2p$^5$$n\ell$ ($n \le$ 7, but for $n$ = 6 and 7, $\ell \le$ 2), and 2s2p$^6$$n\ell$ ($n \le$ 7, but for $n$ = 6 and 7, $\ell \le$ 2) configurations.   For their calculations, they have adopted the general-purpose relativistic atomic structure package (GRASP0 version of P.H.~Norrington and I.P.~Grant)  and the flexible atomic code (FAC), which are both available on the websites \\{\tt http://amdpp.phys.strath.ac.uk/UK\_APAP/codes.html} and {\tt https://www-amdis.iaea.org/FAC/}, respectively.  By performing and comparing the two sets of calculations, they have assessed their energy levels to be accurate to $\sim$0.5~Ryd. However,  some of the levels (particularly the higher ones) between the two calculations differ by up to  $\sim$2~Ryd. In our long experience for a wide range of ions, such a large difference in energy levels with these two different codes (i.e. GRASP and FAC) has normally not been found, and therefore we have performed our own calculations with the same configurations, as adopted by them. Unfortunately, we note that some of their results with FAC cannot be reproduced, and hence the large discrepancies.  In addition, the above listed 209 levels  are not the lowest, and hence there is scope for improvement, particularly for the lifetimes, because some of the neglected levels from other configurations, such as 2p$^5$6f/g/h and 2p$^5$7f/g/h/i,  intermix with these and hence contribute to the calculations.

\section {Energy levels}

Singh et al. \cite{sam} have performed two sets of calculations, with GRASP and FAC, and have included CI (configuration interaction) among 64 configurations, namely 2s$^2$2p$^6$, 2s$^2$2p$^5$$n\ell$ ($n \le$ 7, but for $n$ = 6 and 7, $\ell \le$ 2), 2s2p$^6$$n\ell$ ($n \le$ 7, but for $n$ = 6 and 7, $\ell \le$ 2), 2s$^2$2p$^4$3$\ell$3$\ell'$, 2s$^2$2p$^4$3$\ell$4$\ell'$, and 2s$^2$2p$^4$3$\ell$5$\ell'$. These configurations generate 3948 levels in total, but the results have been reported for only among 209 levels of the above listed lowest 31 configurations alone. However, some of the energies obtained  between the two calculations differ by up to $\sim$2~Ryd, see for example the (2p$^5$) 7p and 7d levels of Ta~LXIV, W~LXV, and Re~LXVI in their Tables~2--4, or present Table~1. In the absence of measurements and other theoretical results it becomes difficult to know `which set of data is more accurate'. Therefore, we have performed calculations with both codes to verify their reported results as well as to further assess the accuracy of the energy levels.

In our calculations with GRASP, we have included the same 3948 levels from 64 configurations, as by Singh et al. \cite{sam}.  However, with FAC we have performed two calculations, i.e. one (FAC1) with the same configurations as with GRASP, and the other  (FAC2) much larger with 6619 levels generated by 2s$^i$2p$^j$ ($i$+$j$ = 8), (2s$^i$2p$^j$, $i$+$j$ = 7) 3$\ell$, 4$\ell$, 5$\ell$, 6$\ell$, 7$\ell$, and (2s$^i$2p$^j$, $i$+$j$ = 6) 3$\ell$4$\ell$, 3$\ell$5$\ell$, and 3$\ell$6$\ell$   configurations. This is to assess the effect of higher lying levels on the accuracy of lower level energies. Results obtained from these three calculations, along with those reported by Singh et al. are listed in Table~1 for a ready comparison, but only for the highest four levels of (2p$^5$) 7p and 7d each, for which the discrepancies are the maximum. 

There are no appreciable differences between our calculations with GRASP and those of Singh et al. \cite{sam} for the levels of Ne-like ions under discussion. However, there are some occasional minor differences for a few levels -- see for example, the 7p~$^3$P$_0$ level of W~LXV in Table~1c. The same is unfortunately not true for the calculations with FAC, because differences for the levels shown in Table~1 are up to 1.8~Ryd. The reason/s for these difference/s are hard to speculate, particularly when the corresponding discrepancies with the lower levels are not as striking. 


Finally, a comparison between our FAC1 and FAC2 energies indicates that there is no appreciable impact of larger CI on the 209 levels of Ne-like ions, because the two sets of calculations agree within $\sim$0.05~Ryd. Therefore, for these levels the CI included in the GRASP1 and FAC1 calculations is sufficient to produce accurate energy levels. Furthermore, as expected, both calculations produce comparable results for a majority of levels, and the differences (if any) are within 0.25~Ryd. This is in contrast to what Singh et al. \cite{sam} have shown, mainly because not only (some of) their results with FAC are incorrect but can also be not reproduced. It may be worth mentioning here that we have noted similar problems in the past with their calculations with both the GRASP and FAC codes -- see for example, the energy levels of five Br-like ions \cite{brlike} with 38 $\le$ Z $\le$ 42 and F-like W~LXVI \cite{w66a},\cite{w66b}. 

\begin{table}
\caption{Comparison of some energy levels of Ne-like ions.} 
\begin{tabular}{lllllllll}  \hline
Level (2p$^5$) & 7p~$^3$D$_1$ & 7p~$^3$P$_0$ & 7p~$^3$S$_1$ & 7p~$^1$D$_2$ & 7d~$^3$F$^o_2$ & 7d~$^3$D$^o_1$ & 7d~$^3$P$^o_2$ & 7d~$^1$F$^o_3$ \\
 \hline
 {\bf a.} Hf~LXIII\\
 GRASP1a & 1083.665 & 1083.837 & 1085.502 & 1085.506 & 1086.274 & 1086.377 & 1086.750 & 1086.762 \\
 GRASP1b & 1083.651 & 1083.821 & 1085.489 & 1085.493 & 1086.261 & 1086.364 & 1086.737 & 1086.749 \\
 FAC1a   & 1083.972 & 1084.129 & 1084.972 & 1085.168 & 1085.835 & 1085.841 & 1086.250 & 1086.259 \\
 FAC1b   & 1083.707 & 1083.861 & 1085.554 & 1085.567 & 1086.350 & 1086.450 & 1086.831 & 1086.846 \\
 FAC2    & 1083.658 & 1083.815 & 1085.526 & 1085.520 & 1086.303 & 1086.411 & 1086.781 & 1086.798 \\
 \hline
 {\bf b.} Ta~LXIV \\
 GRASP1a & 1120.443 & 1120.620 & 1222.404 & 1122.407 & 1123.189 & 1123.293 & 1123.695 & 1123.707 \\
 GRASP1b & 1120.429 & 1120.603 & 1122.391 & 1122.393 & 1123.175 & 1123.279 & 1123.682 & 1123.693 \\
 FAC1a   & 1120.758 & 1120.918 & 1121.396 & 1121.590 & 1122.694 & 1122.707 & 1122.745 & 1122.745 \\
 FAC1b   & 1120.486 & 1120.644 & 1122.456 & 1122.469 & 1123.267 & 1123.368 & 1123.778 & 1123.793 \\
 FAC2    & 1120.436 & 1120.597 & 1122.422 & 1122.422 & 1123.219 & 1123.324 & 1123.728 & 1123.745 \\
 \hline
 {\bf c.} W~LXV \\
 GRASP1a & 1158.024 & 1158.205 & 1160.113 & 1160.116 & 1160.912 & 1161.017 & 1161.450 & 1161.462 \\
 GRASP1b & 1158.009 & 1158.187 & 1160.099 & 1160.102 & 1160.897 & 1161.003 & 1161.436 & 1161.448 \\
 FAC1a   & 1158.343 & 1158.509 & 1158.609 & 1158.803 & 1159.928 & 1159.941 & 1160.461 & 1160.464 \\
 FAC1b   & 1158.067 & 1158.228 & 1160.167 & 1160.180 & 1160.991 & 1161.094 & 1161.535 & 1161.551 \\
 FAC2    & 1158.013 & 1158.179 & 1160.130 & 1160.133 & 1160.943 & 1161.049 & 1161.485 & 1161.503 \\
 \hline
 {\bf d.} Re~LXVI\\
 GRASP1a & 1196.421 & 1196.606 & 1198.647 & 1198.650 & 1199.458 & 1199.565 & 1200.031 & 1200.042 \\
 GRASP1b & 1196.406 & 1196.588 & 1198.633 & 1198.635 & 1199.443 & 1199.551 & 1200.016 & 1200.028 \\
 FAC1a   & 1196.608 & 1196.768 & 1196.807 & 1196.934 & 1197.972 & 1197.985 & 1199.002 & 1199.008 \\
 FAC1b   & 1196.465 & 1196.629 & 1198.702 & 1198.715 & 1199.540 & 1199.644 & 1200.118 & 1200.134 \\
 FAC2    & 1196.430 & 1196.596 & 1198.663 & 1198.668 & 1199.492 & 1199.598 & 1200.067 & 1200.086 \\
\hline	
\end{tabular}

\begin{flushleft}
{\small
GRASP1a: calculations of Singh et al. \cite{sam} with GRASP for 3948 levels \\
GRASP1b: present calculations with GRASP for 3948 levels \\
FAC1a: calculations of Singh et al. with FAC for 3948 levels \\
FAC1b: present calculations with FAC for 3948 levels \\
FAC2: present calculations with FAC for 6619 levels \\

}
\end{flushleft}
\end{table}

As discussed above, there is not much  scope for improvement in the calculations of energy levels for the four Ne-like ions considered here. However, the corresponding calculations for lifetimes ($\tau$) can certainly be improved, because levels from some of the neglected configurations, such as 2p$^5$6f/g/h and 2p$^5$7f/g/h/i,  intermix with these and hence contribute to the determination of $\tau$. Similarly, the limited results for radiative rates (A-values) reported by Singh et al. \cite{sam}, mainly from the ground to higher excited levels, are insufficient for the accurate modelling of plasmas, because a complete set of data for {\em all} transitions is often required.

\section{Conclusions}

Recently, Singh et al. \cite{sam} reported results for  energy levels, A-values, and $\tau$  among 209 levels of four Ne-like ions with  72 $\le$ Z $\le$ 75. Particularly for energy levels they performed two sets of calculations with the GRASP and FAC codes. This was to assess the accuracy of energies, because prior similar data, experimental or theoretical, are almost non-existent, except for W~LXV. For many levels their two sets of energies differ by $\sim$0.5~Ryd, but for a few (particularly the higher ones) the discrepancies are up to $\sim$2~Ryd. This is in spite of adopting the same level of CI and including the contribution of relativistic effects in both calculations. Since such large differences between any two independent calculations have neither been noted earlier nor are expected, we have performed fresh calculations with the same two codes and with the same level of CI. On the basis of detailed comparisons made, among our various calculations as well as with the work of Singh et al., our conclusion is that there is no (appreciable) discrepancy for any level between the energies obtained with GRASP and FAC. Conversely, some of the level energies reported by Singh et al., with the FAC code, are incorrect and cannot be reproduced.

With an inclusion of even larger CI, than considered by Singh et al. \cite{sam}, there is no significant change, either in magnitude or orderings, for the 209 levels of Ne-like ions, which belong to the 2s$^2$2p$^6$, 2s$^2$2p$^5$$n\ell$ ($n \le$ 7, but for $n$ = 6 and 7, $\ell \le$ 2), and 2s2p$^6$$n\ell$ ($n \le$ 7, but for $n$ = 6 and 7, $\ell \le$ 2) configurations. However,    some levels of higher neglected configurations, such as (2p$^5$)~6f/g/h and 7f/g/h/i,  intermix with these and hence their A-values contribute to the determination of $\tau$. Therefore, there is scope for improvement over the calculations of $\tau$ for some (about a third of) levels, i.e.  higher than 137 -- see Tables~1--4 of \cite{sam}. Similarly, the limited results of A-values reported by Singh et al. are insufficient for a reliable plasma modelling, for which a complete set of data is preferably required.  A complete set of energies and A-values for three Ne-like ions, namely Hf~LXIII, Ta~LXIV and Re~LXVI  are reported in our recent paper \cite{nelike}, whereas for W~LXV in an earlier one \cite{w65}.


\end{document}